\documentclass[aps,pra,twocolumn,groupedaddress,reprint,longbibliography]{revtex4-2}

\usepackage{graphicx}
\usepackage{epstopdf}
\usepackage{comment}
\usepackage{amsmath,amssymb}
\usepackage{bm}
\usepackage{float}
\usepackage[utf8]{inputenc}
\usepackage[T1]{fontenc}
\usepackage{color}

\newcommand{\ctext}[1]{\raise0.2ex\hbox{\textcircled{\scriptsize{#1}}}}
\newcommand{\vect}[1]{{\mbox{\boldmath $#1$}}}

\begin{document}

\title{Multimode oscillation and its route to chaos \\in transverse-mode-coupled optomechanical resonators}

\author{Motoki Asano}
\email[]{motoki.asano@ntt.com}
\affiliation{Basic Research Laboratories, NTT, Inc., Kanagawa, Japan}

\author{Lalit Maharjan}
\affiliation{Basic Research Laboratories, NTT, Inc., Kanagawa, Japan}

\author{Hiroshi Yamaguchi}
\affiliation{Basic Research Laboratories, NTT, Inc., Kanagawa, Japan}

\date{\today}

\begin{abstract}
Nonlinear oscillators exhibit rich collective phenomena including synchronization, bifurcation, and chaos. Experimentally accessing multimode nonlinear dynamics, however, remains challenging because low-order internal resonances readily induce mode locking and suppress stable multimode oscillation. Here, we demonstrate a transverse-mode-coupled optomechanical platform that combines engineered collective mechanical modes with optically controllable nonlinear interactions. The engineered transverse mechanical modes enable stable two- and three-mode self-sustained oscillations while suppressing trivial low-order internal resonances. By varying the laser frequency, we observe successive beat-frequency fractionalization (\(\Delta f/2\) and \(\Delta f/3\)), followed by a broadband state exhibiting enhanced trajectory divergence consistent with deterministic chaos.  These results establish transverse-mode-coupled optomechanical resonators as a scalable and tunable platform for exploring high-dimensional nonlinear dynamics in interacting mechanical networks.
\end{abstract}

\maketitle

\section{Introduction}
Interactions among many nonlinear oscillators give rise to emergent dynamical behaviors that cannot be understood from isolated oscillators alone. Among these, multimode oscillatory states, in which several independent frequencies coexist simultaneously, provide a fundamental platform for studying synchronization, quasiperiodicity, bifurcation cascades, and chaos. Understanding how such multimode states emerge, evolve, and eventually lose stability is therefore a central problem in nonlinear dynamics, with implications ranging from physical and biological systems to bio-inspired dynamical architectures for information processing.

Micro- and nanoelectromechanical resonators provide a well-established experimental platform for investigating nonlinear phenomena. Their higher-order mechanical modes, together with structural frequency engineering, allow modal frequencies to be brought close to simple commensurate ratios such as 1:2 and 1:3 \cite{antonio2012frequency,eichler2012strong,guttinger2017energy,houri2020demonstration,kecskekler2022symmetry,macdonald2016optomechanics,matheny2013nonlinear,qalandar2014frequency,shoshani2017anomalous}. The resulting low-order internal resonances selectively enhance nonlinear coupling pathways and have been widely utilized to achieve functionalities such as coherent nonlinear energy transfer and mechanical frequency-comb generation \cite{chen2017direct, czaplewski2018bifurcation, fu2025sideband}. Such resonant interactions, however, often impose phase-locking conditions among the participating modes that constrain their independent degrees of freedom \cite{arnold1989mathematical}.

Away from dominant low-order internal resonances, multiple independent oscillatory degrees of freedom can be preserved, providing access to high-dimensional nonlinear dynamics characterized by multidimensional invariant tori, resonance networks, and routes to high-dimensional chaos. Such high-dimensional chaotic dynamics are intimately connected to complex phenomena including turbulence and spatiotemporal chaos \cite{ruelle1971nature, cross1993pattern,grebogi1983three,baesens1991three,yamagishi2020chaos}, making their implementation in a well-controlled physical platform particularly valuable for experimentally exploring the emergence and control of complex dynamics \cite{gollub1980many,cumming1988quasiperiodicity,winful1986frequency}. For this purpose, mechanical systems readily support scalable arrays of coupled modes \cite{buks2002electrically,sato2003observation,asano2024fiber,ren2022topological}. The central challenge is to preserve multiple independent oscillatory degrees of freedom as the number of interacting modes increases by simultaneously avoiding dominant low-order resonances and sustaining multimode self-oscillation.

In this work, we demonstrate such a platform using an array of elastically coupled optomechanical resonators. Elastic coupling splits the mechanical resonance into a set of spectrally resolved collective modes, hereafter referred to as transverse mechanical modes. By engineering their frequency spectrum, we suppress dominant low-order internal resonances while enabling stable self-sustained oscillation of multiple modes. A single optical cavity driven by a continuous-wave laser provides shared mechanical gain and nonlinear coupling, allowing two and three transverse modes to oscillate simultaneously without separate actuators or mode-specific feedback. By tuning the laser frequency and input optical power, we experimentally observe transitions from stable multimode oscillation through successive beat-frequency fractionalization and ultimately to deterministic chaos, providing experimental achievement of the rich bifurcation dynamics recently predicted for multimode phonon lasing \cite{ortiz2025dual}. These results establish transverse-mode engineering combined with shared optomechanical gain as a scalable route to complex multimode dynamics beyond dominant low-order internal resonances.

\begin{figure}[htbp]
\centering
\includegraphics[width=\linewidth]{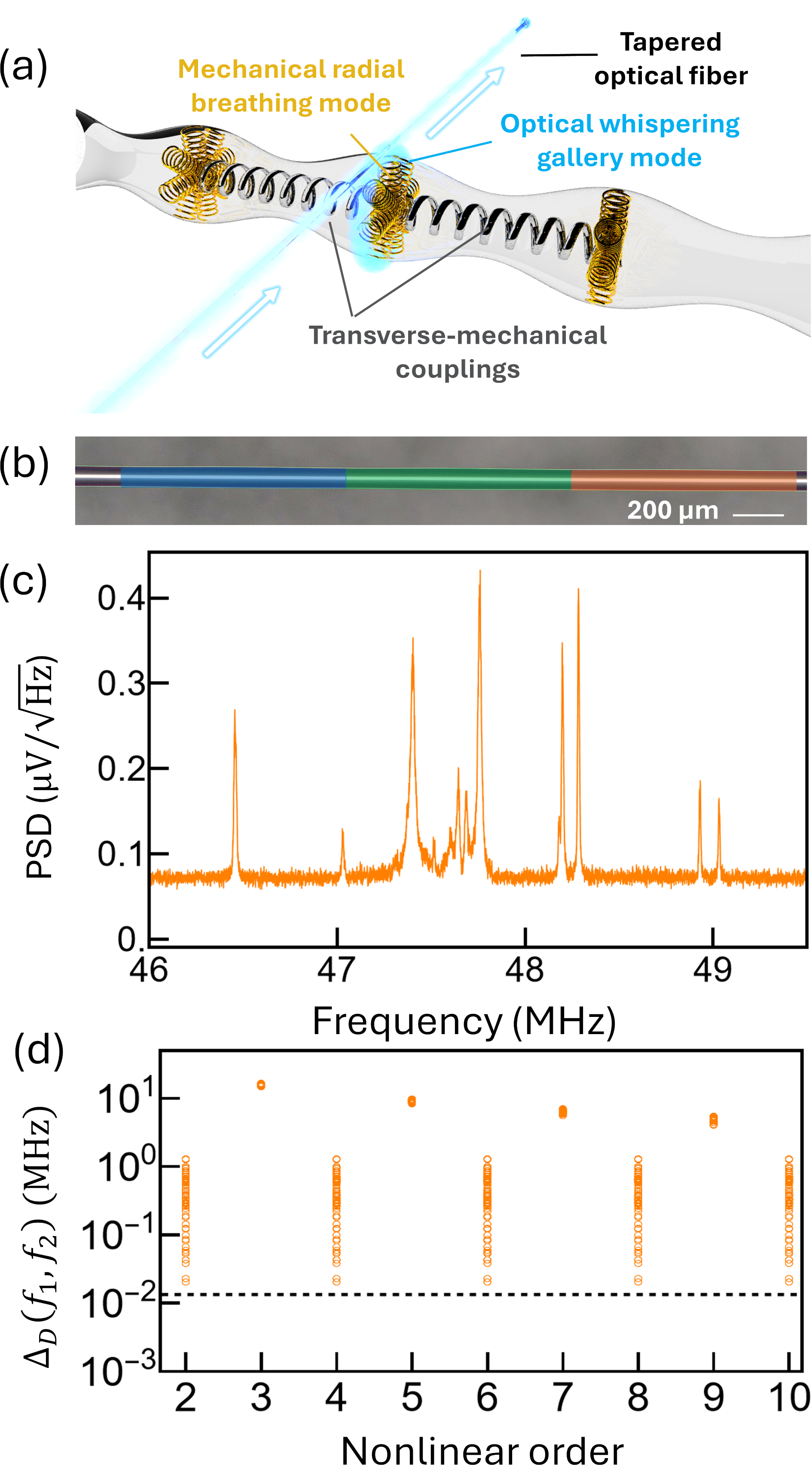}
\caption{(a) Concept of coupled microbottle optomechanical array. (b) Optical microscope image of fabricated device. Differently colored shaded areas indicate individual microbottle structures. (c) Thermally driven mechanical fluctuation spectrum showing multiple collective mechanical modes. (d) Minimum resonance detuning, $\Delta_D(\vect{f})$, for different nonlinear order $D$. Dashed line denotes average mechanical linewidth, indicating that low-order internal resonances remain spectrally resolved.}
\label{fig:1}
\end{figure}
\section{Experiment}
Figure 1(a) illustrates the concept of the transverse-mode-coupled optomechanical resonator based on a microbottle resonator array. In an individual microbottle resonator, higher-order radial breathing modes form a longitudinal mode spectrum with frequencies approaching simple harmonic relations. By contrast, arranging resonators along the fiber axis introduces a set of collective transverse mechanical modes whose frequencies remain clustered around a common radial breathing resonance. Elastic coupling further splits these modes into spectrally resolved collective states, and the number of accessible transverse modes scales naturally with the number of resonators \cite{asano2022free,asano2024fiber}.

Figure~1(b) shows an optical microscope image of the optomechanical array that comprises three microbottle structures. This resonator was fabricated on an 80-$\mathrm{\mu m}$-diameter silica fiber using the heat-and-pull technique \cite{asano2016observation}. An optical tapered fiber was brought into contact with a microbottle resonator to excite whispering-gallery-modes (WGMs) in the wavelength range from 1530~nm to 1570~nm. The optical quality factor is typically in the order of \(10^7\), comparable to those of previously reported microbottle resonators \cite{asano2016observation,macdonald2016optomechanics,asano2025synthesized}.

To characterize the multiple mechanical modes, we first measured thermally driven mechanical fluctuations by stabilizing a continuous-wave laser on the slope of an optical resonance with 10-mW optical power. The details of the experimental setup are provided in Appendix A. Figure~1(c) shows the power spectral density of the thermal fluctuations. Multiple mechanical resonances were observed between 46 MHz and 50 MHz, originating from the hybridization of multiple higher-order axial modes of the three coupled resonators \cite{asano2024fiber}.

Here we define the minimum resonance detuning for a set of eigenfrequencies, $\vect{f}$, to quantify the proximity to low-order internal resonances as
\begin{equation}
\Delta_D(\vect{f})=
\min_{\sum_i |n_i|=D}
\frac{\left|\sum_i n_i f_i\right|}{D}.
\end{equation}
Quantity $\Delta_D(\vect{f})$ represents the smallest normalized frequency mismatch among resonance conditions of nonlinear order $D$. Figure~1(d) summarizes the distributions of $\Delta_D$ calculated from the experimentally observed eigenfrequencies for different nonlinear orders. The stepwise dependence of $\Delta_D$, with identical values for successive orders, arises because the closest rational approximants for the nearly degenerate modes ($f_1 \simeq f_2$) predominantly involve $|n_1| \simeq |n_2|$ and hence even values of $D$. As a reference scale for the spectral resolution of internal resonances, the dashed line indicates the average mechanical linewidth extracted from the thermal fluctuation spectra. Even for nonlinear orders up to $D=10$, the calculated resonance detunings remain well above the mechanical linewidth, indicating that low-order internal resonances are spectrally well resolved and therefore unlikely to induce trivial resonance locking.

The evolution of the oscillation dynamics is examined by varying the laser wavelength at a fixed input optical power of 120~mW. Figures~2(a) and 2(b) show optical transmittance and a map of the mechanical power spectral density, respectively. Over a broad wavelength range, single-mode self-sustained oscillations with a phonon linewidth below the measurement resolution are observed, consistent with stable limit-cycle oscillation [see Fig.~2(c)]. As the laser wavelength is swept across different optical resonances, the oscillating mechanical mode switches between several mechanical modes. Such mode selection is consistent with competition governed by the balance among spatially dependent optomechanical gain, photothermal interactions, and optically induced nonlinear damping \cite{kemiktarak2014mode}. In certain wavelength regions, however, this competition is relaxed and two mechanical modes oscillate simultaneously [Fig.~2(d)]. This produces nonlinear mixing components at integer combinations of their oscillation frequencies and thereby generates evenly spaced comb-like features in the measured spectrum. This stable two-mode oscillation is consistent with a balance between modal gain and nonlinear saturation, including optically induced nonlinear damping \cite{arahari2026optically}. Moreover, the parameter region supporting stable two-mode oscillation expands with increasing input optical power, as described in Appendix B.

To characterize the resonance conditions of the observed oscillation states, we repeat the wavelength sweep at different input optical powers and calculate $\Delta_{10}(\vect{f})$ from the oscillation frequencies identified in each spectrum. Figure~2(e) shows the mean value of $\Delta_{10}(\vect{f})$ at each input optical power, with the error bars representing the standard deviation over the corresponding wavelength sweep. Throughout the investigated power range, $\Delta_{10}(\vect{f})$ remains well above the mechanical linewidth, showing that the observed oscillations are spectrally separated from low-order internal resonances.

\begin{figure}[htbp]
\centering
\includegraphics[width=\linewidth]{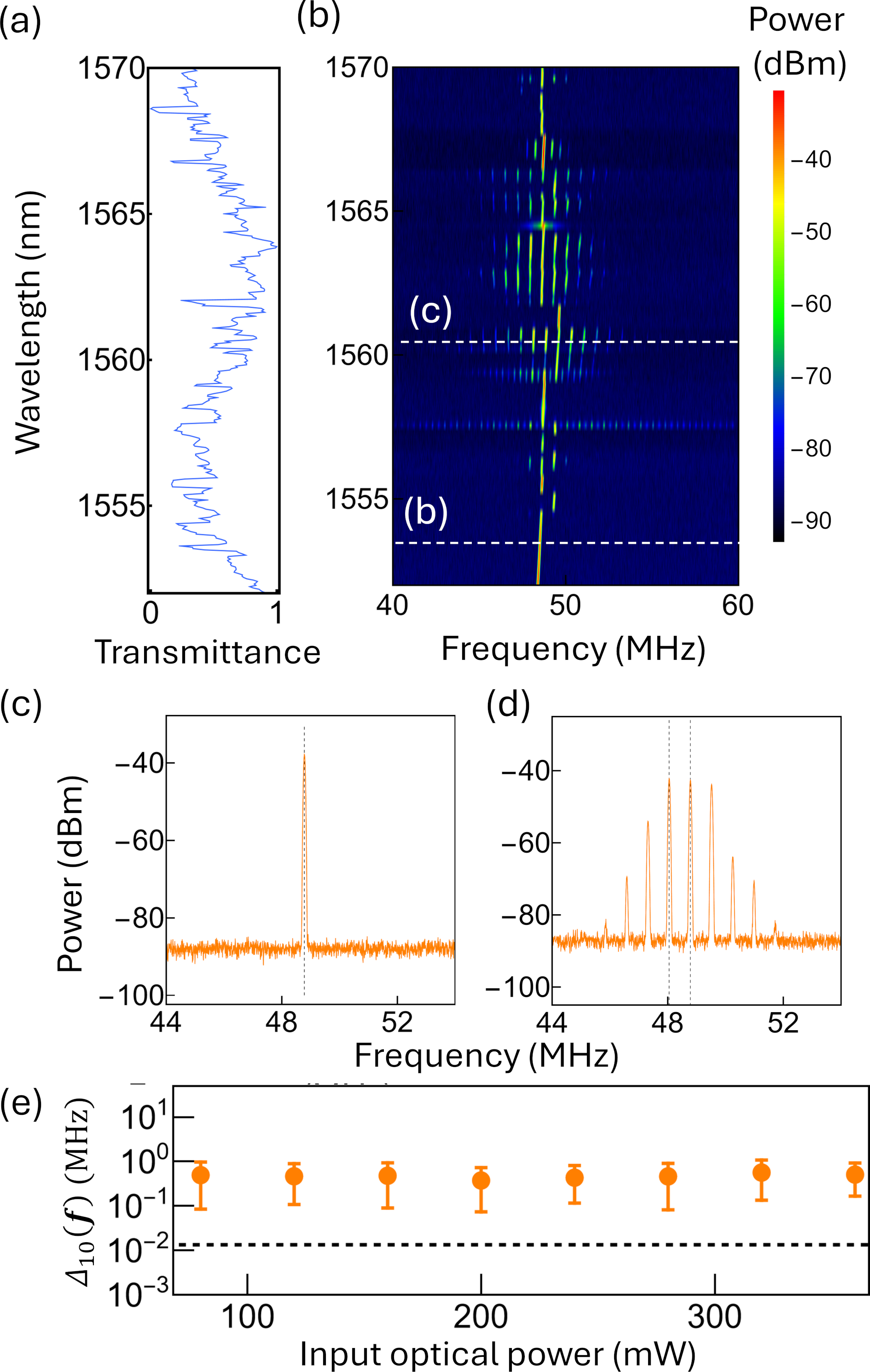}
\caption{(a) Optical transmission and (b) mechanical power spectra measured while sweeping the laser wavelength. (c) Representative single-mode self-sustained oscillation. (d) Representative two-mode self-sustained oscillation with nonlinear mixing sidebands. (e) Average minimum resonance detuning $\Delta_{10}(\mathbf{f})$ as a function of input optical power. Error bars indicate the standard deviation, and the dashed line denotes the average mechanical linewidth.
}
\label{fig:2}
\end{figure}

In addition to two-mode oscillation, we identify operating conditions supporting self-sustained oscillation of three distinct mechanical modes. Figure~3(a) shows a representative three-mode state observed at an input optical power of 380~mW and a laser wavelength of 1568~nm. Three narrow spectral peaks corresponding to the mechanical eigenmodes at $f_1$, $f_2$, and $f_3$ are simultaneously observed [Fig.~3(b)], together with multiple nonlinear mixing components at integer combinations of these frequencies. The narrow linewidths of the three principal peaks indicate a stable three-mode oscillatory state.

To assess the proximity of the observed three-mode states to internal resonances, we evaluate $\Delta_D(\vect{f})$ for the three oscillation frequencies under two distinct optical operating conditions. As shown in Fig.~3(c), $\Delta_D(\vect{f})$ remains well above the characteristic mechanical linewidth at low nonlinear orders for both conditions, indicating that the observed three-mode oscillations are not stabilized by trivial low-order internal resonance locking. Although $\Delta_D(f)$ generally decreases with increasing nonlinear order because of the growing number of possible near-resonant combinations, the observed three-mode state remains spectrally separated from internal resonances over the low-order range relevant here. These observations show that the engineered collective-mode spectrum supports stable multimode oscillation without relying on dominant low-order internal resonances. To the best of our knowledge, simultaneous optomechanical self-oscillation involving three or more mechanical modes has previously been reported only in a system assisted by self-pulsing dynamics arising from coupled free-carrier and thermal effects \cite{alonso2026multimode}. By contrast, the present three-mode oscillation emerges directly from optomechanical gain and intermodal interactions under continuous-wave optical driving, without relying on an additional self-pulsing degree of freedom.

\begin{figure}[t]
\centering
\includegraphics[width=\linewidth]{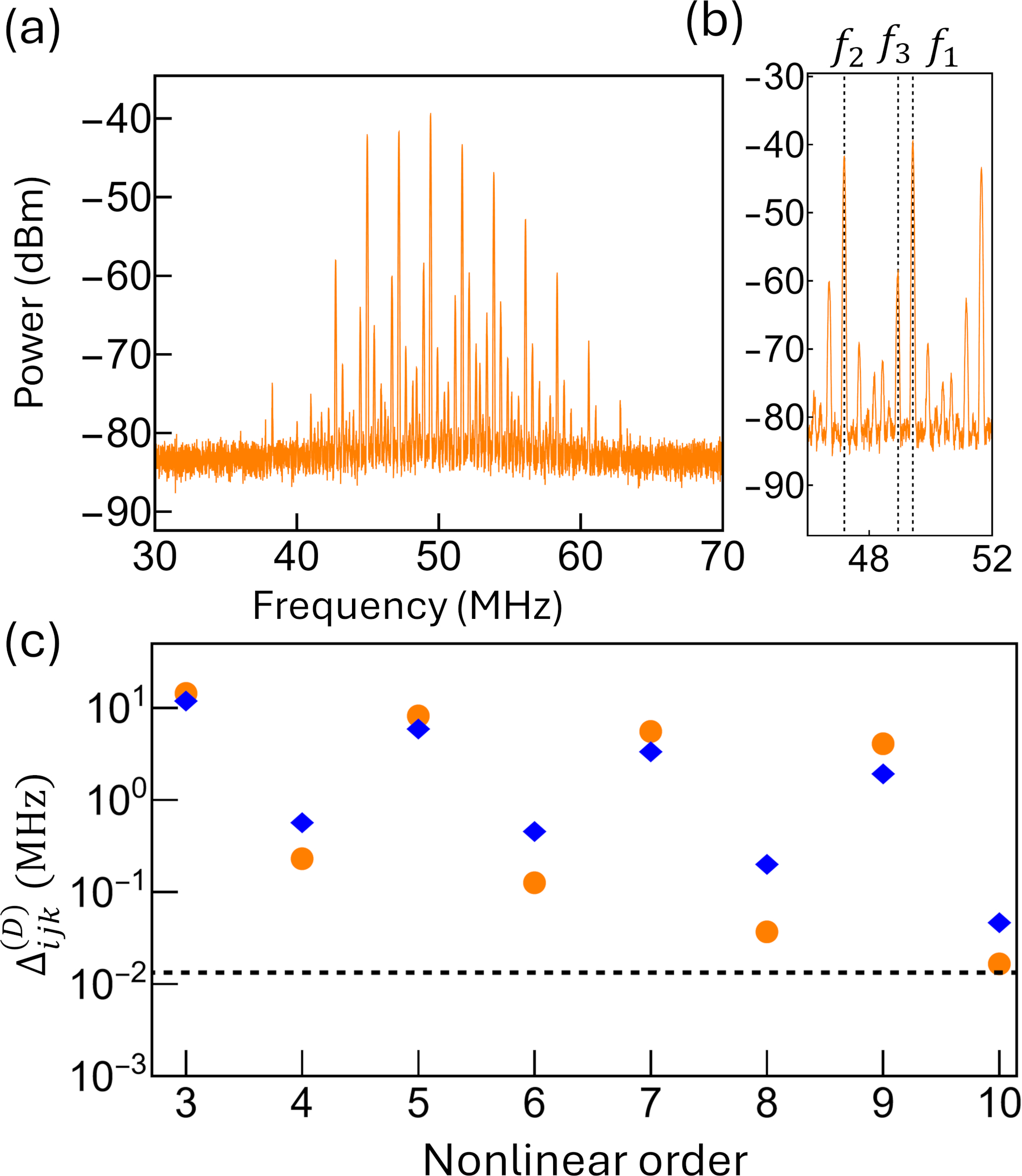}
\caption{(a) Representative mechanical power spectrum showing simultaneous oscillation of three collective modes. (b) Enlarged view of the three oscillation peaks. (c) Minimum resonance detuning $\Delta_D(\mathbf{f})$ evaluated under two optical operating conditions. Dashed line indicates average mechanical linewidth.
}
\label{fig:3}

\end{figure}
\begin{figure}[htpb]
\centering
\includegraphics[width=\linewidth]{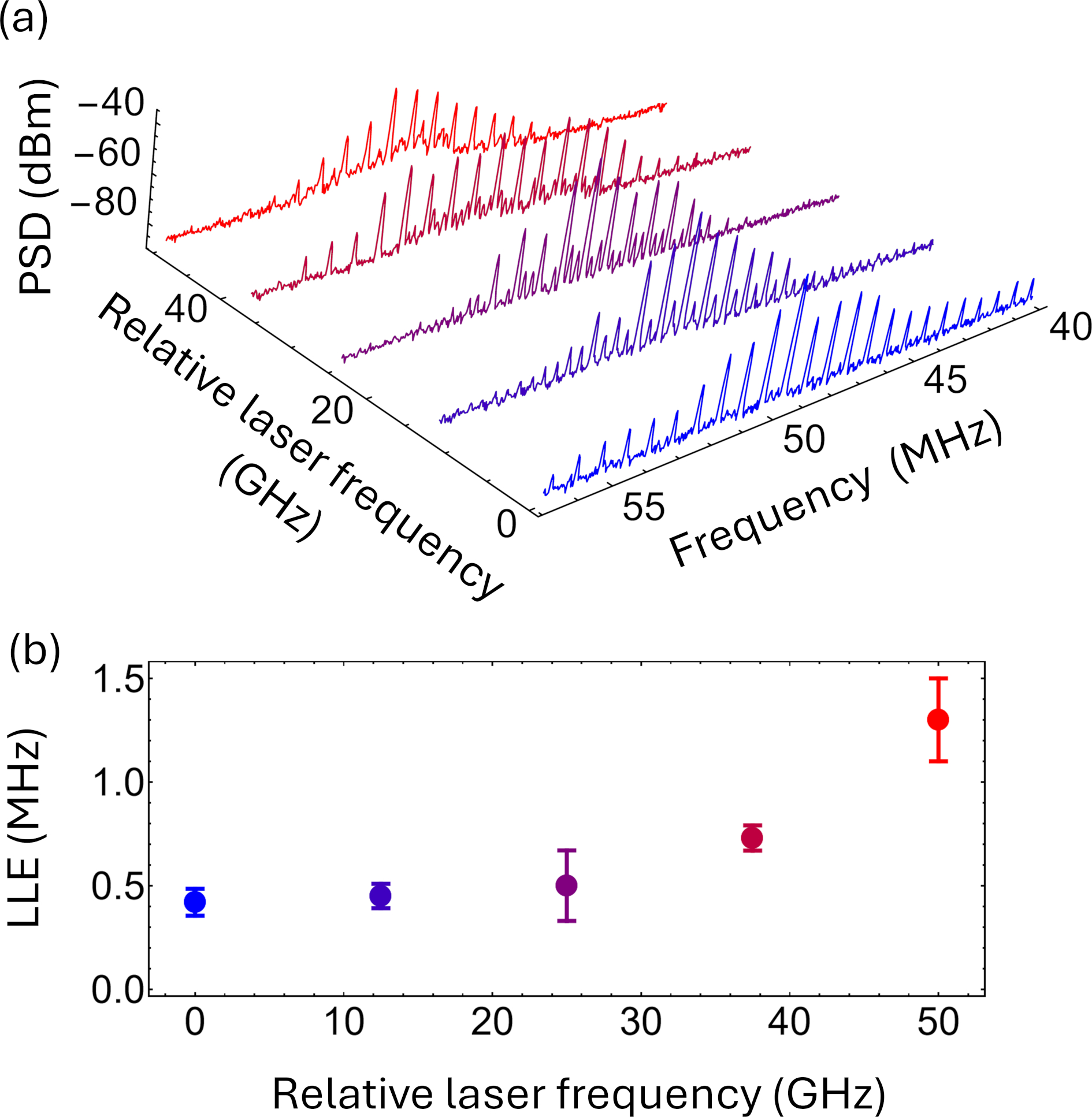}
\caption{(a) Mechanical power spectra as laser frequency is varied, showing successive emergence of $\Delta f/2$ and $\Delta f/3$ beat-frequency components before transition to a broadband state. (b) Estimated largest Lyapunov exponent (LLE) obtained from experimentally measured time series.
}
\label{fig:4}
\end{figure}

The stable multimode oscillatory states established above provide an ideal starting point for exploring higher-order nonlinear dynamics. We therefore investigate how a stable two-mode oscillatory state evolves as the laser frequency is varied. Figure~4(a) summarizes the corresponding mechanical power spectra. The spectrum comprises discrete components separated by the intermodal beat frequency $\Delta f=f_2-f_1$, which is characteristic of stable two-mode oscillation. As the laser frequency increases, additional spectral components with spacings of $\Delta f/2$ and subsequently $\Delta f/3$ emerge before the transition to a broadband state.

The appearance of these fractional beat-frequency components indicates a nonlinear restructuring of the intermodal dynamics. Recent bifurcation analysis of dual-mode optomechanical oscillators has predicted transitions involving toroidal dynamics, subharmonic generation, and deterministic chaos \cite{ortiz2025dual}. The observed sequence from $\Delta f$ to $\Delta f/2$, $\Delta f/3$, and finally the broadband state is therefore consistent with a bifurcation-mediated route toward chaos. The same evolution from the $\Delta f/3$ state to a broadband state is reproduced at another optical resonance, as described in Appendix C, indicating that the observed transition is not restricted to a single optical operating condition.

To characterize further the broadband state, we estimated the largest Lyapunov exponent (LLE) from the experimentally measured time series using the Rosenstein algorithm \cite{rosenstein1993practical}. As shown in Fig. 4(b), the estimated LLE exhibits a small positive residual value in the regular oscillatory regime, but increases markedly toward the broadband state. The largest value is obtained in the broadband regime, indicating substantially enhanced trajectory divergence. Together with the successive emergence of fractional beat-frequency components and the broadband spectrum, this pronounced enhancement of the estimated LLE provides evidence that the broadband state corresponds to deterministic chaos. A similar enhancement is reproduced at a different optical resonance (Appendix C), showing that the observed behavior is robust across different optical operating conditions.

\section{Discussion}
The present microbottle platform combines scalability with optical control of nonlinear mechanical dynamics. Elastically coupled microbottle resonator arrays have previously been extended to as many as 50 resonators \cite{asano2024fiber}, establishing that the underlying coupling architecture is not restricted to a few-element system. In addition to this structural scalability, previous studies have established optical control over mechanical gain, Duffing nonlinearity, and nonlinear damping in the same platform \cite{asano2025synthesized,arahari2026optically}. The resulting system therefore provides a scalable and optically tunable framework for engineering collective nonlinear mechanical dynamics. Although the present experiments show simultaneous oscillation of up to three mechanical modes, excitation through a larger number of optical modes or independently controlled optical pumps may extend the accessible regime toward higher-dimensional multimode oscillation. Such an extension will, however, require control over the increasingly dense network of modal competition and near-resonant nonlinear interaction channels.

A distinctive feature of the observed chaos is that the bifurcation develops in the intermodal beat dynamics between closely spaced mechanical modes. Multimode phonon lasing and nonlinear dynamics have also been explored for mechanical modes with widely separated frequencies \cite{ng2023intermodulation,wang2024optomechanical,ortiz2025dual}. By contrast, the present approach utilizes multiple closely spaced mechanical modes, whose small frequency differences introduce slow intermodal degrees of freedom on top of the much faster carrier oscillations. The successive appearance of spectral spacings at $\Delta f/2$ and $\Delta f/3$ indicates a progressive restructuring of these slow intermodal dynamics preceding the transition to deterministic chaos. Because arrays naturally provide an increasing number of closely spaced collective modes, this mechanism may offer a scalable route toward higher-dimensional nonlinear dynamics based on multiple interacting mechanical degrees of freedom. Combining such intermodal chaos with intramodal nonlinearities, optical chaos, or slower photothermal dynamics could further provide access to chaotic states spanning multiple dynamical time scales.

\section{Conclusion}
In summary, we demonstrated a multimode route to chaos in an optomechanical array by engineering collective transverse mechanical modes through elastic coupling between microbottle resonators. The engineered frequency landscape supports stable self-sustained oscillation involving both two and three mechanical modes while suppressing trivial low-order internal resonances. By tuning the laser frequency, we observed a reproducible sequence of dynamical transitions characterized by the successive emergence of $\Delta f/2$ and $\Delta f/3$ beat-frequency components before the onset of deterministic chaos. Together with the achievement of stable three-mode oscillation, these results establish elastically coupled optomechanical arrays as a scalable and optically tunable platform for investigating high-dimensional nonlinear dynamics in multimode mechanical networks.

\section*{ACKNOWLEDGMENTS}
The authors thank Hideki Aarahari, Yuta Miyasaka, Takuya Ikuta, Kensuke Inaba, Yasuhiro Yamada, and Takahiro Inagaki for their fruitful discussions. The authors also thank Hajime Okamoto for his continuous encouragement. This work was supported by JSPS KAKENHI Grant Numbers JP23H05463 and JP26K01424. 

\bibliography{sample}

\appendix

\section{Experimental setup}
Figure 5 shows the experimental setup used in this work. An external-cavity diode laser (ECDL) was used to excite optical WGMs via a tapered optical fiber brought into contact with the middle microbottle structure in the coupled microbottle resonator. The input optical power was adjusted using an erbium-doped fiber amplifier (EDFA) and a variable optical attenuator (VOA), while the input polarization was controlled using a polarization controller (PC). The transmitted optical signal was detected using a photodetector (PD), and the resulting electrical signal was analyzed using an electrical spectrum analyzer (ESA) and a digital sampling oscilloscope (DSO).
\begin{figure}[htpb]
\centering
\includegraphics[width=\linewidth]{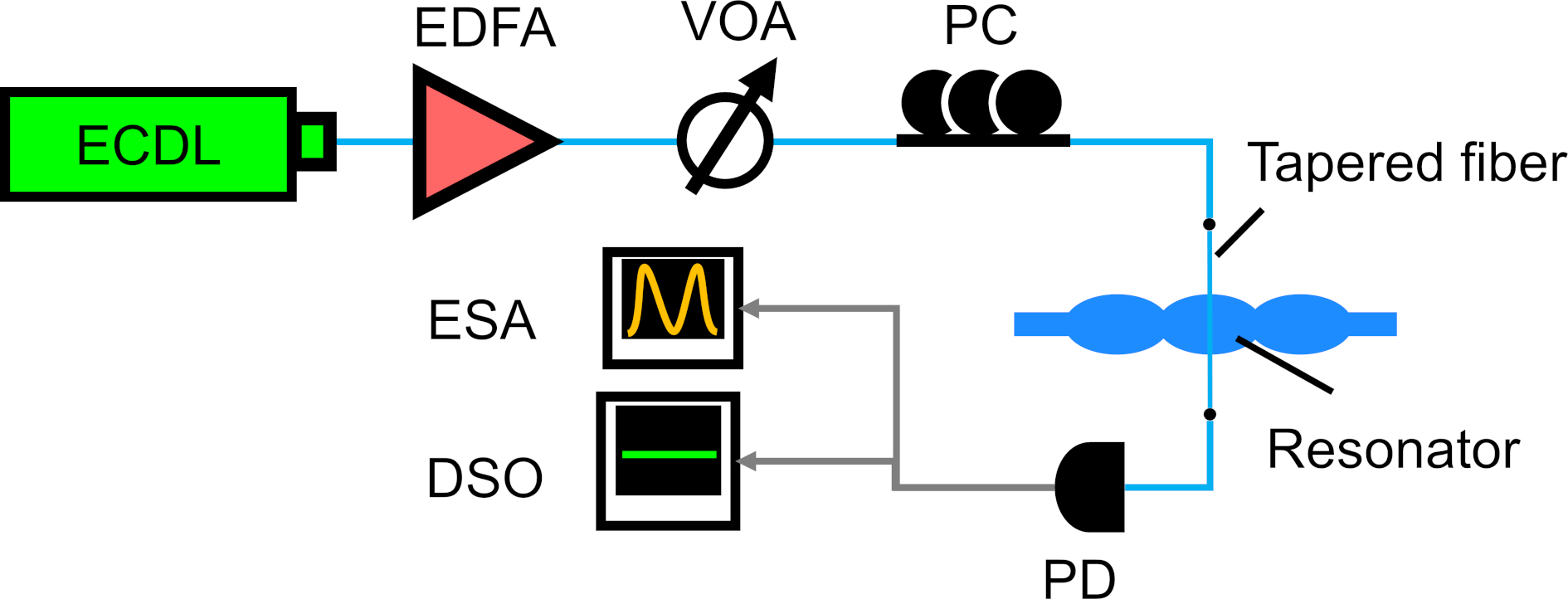}
\caption{Measurement setup}
\label{fig:S1}
\end{figure}

\section{Multimode oscillation spectra}
The multimode oscillation spectra measured at different input optical powers are shown in Fig.~6. As the input optical power $P_{\mathrm{in}}$ is increased, the oscillation evolves from a predominantly single-mode state to a predominantly two-mode state owing to the enhanced mechanical nonlinearities.
\begin{figure*}[htpb]
\centering
\includegraphics[width=\linewidth]{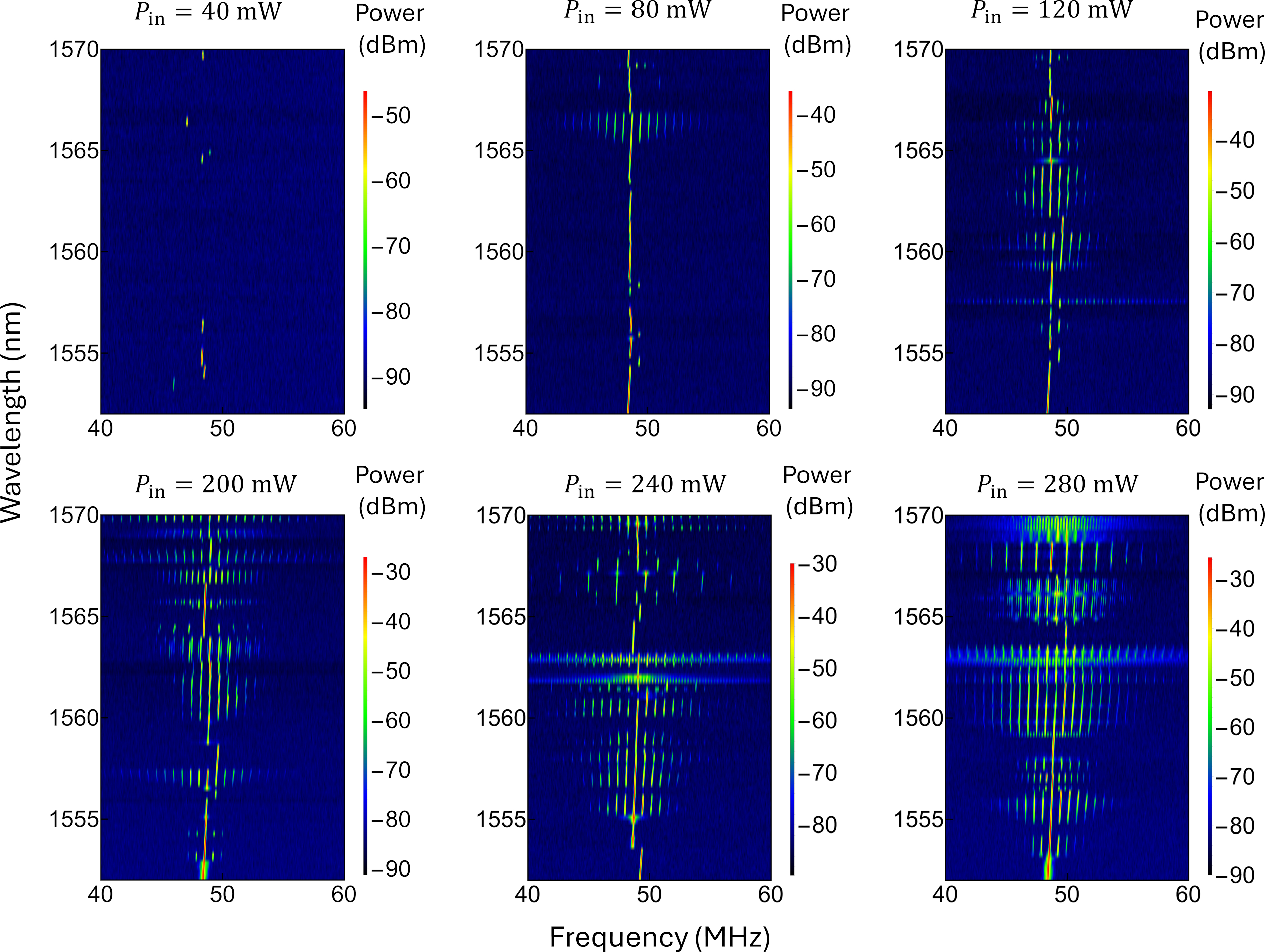}
\caption{Mechanical power spectra measured while sweeping the laser wavelength at different input optical powers.}
\label{fig:S2}
\end{figure*}

\section{Reproducibility of the multimode route to chaos}
To examine the robustness of the observed transition, we performed the same analysis using measurements obtained at a different optical resonance. Figure 7 shows the power spectral density of the mechanical modes and the estimated LLE for representative oscillatory states obtained under this condition. A small positive residual value is again observed for the regular oscillatory states, whereas the broadband state exhibits a substantially larger estimated LLE. Although the absolute LLE values differ from those in Fig. 4(b), as expected for different optical operating conditions, the pronounced enhancement upon entering the broadband regime is reproduced. This result confirms that the observed increase in trajectory divergence is not specific to a particular optical resonance.
\begin{figure}[htpb]
\centering
\includegraphics[width=\linewidth]{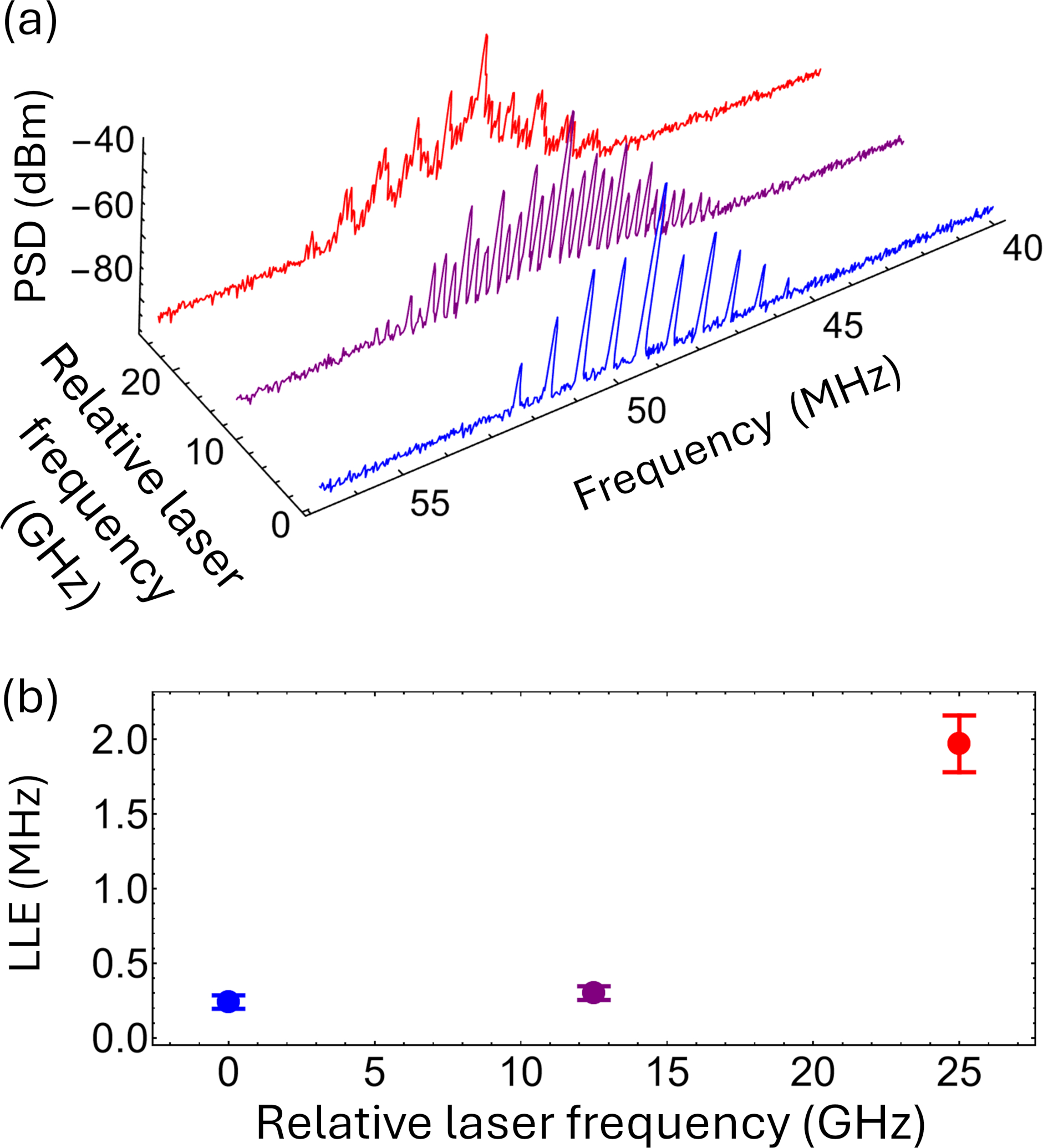}
\caption{
Route from stable two-mode oscillation to deterministic chaos. (a) Mechanical power spectra as the laser frequency is varied. (b) Estimated LLE calculated using the Rosenstein algorithm as a function of laser frequency.
}
\label{fig:4}
\end{figure}

\section{Estimation of the largest Lyapunov exponent}
The LLE was estimated from the experimentally measured time series using the Rosenstein algorithm~\cite{rosenstein1993practical}.
The embedding delay was determined from the autocorrelation of the time series, and the embedding dimension was evaluated using the false-nearest-neighbor method.
The fraction of false nearest neighbors was saturated above $m \simeq 10$; therefore, we used $m=11$.
A Theiler window of 500 samples was used to exclude temporally correlated neighbors, and the logarithmic divergence was averaged over 100 reference trajectories.
The LLE was obtained from the slope of the initial approximately linear region of the mean logarithmic divergence.

To evaluate the uncertainty, each time series was divided into five nonoverlapping temporal segments and the same analysis was independently applied to each segment.
The plotted values and error bars represent the mean and standard deviation of the resulting estimates, respectively.

A small positive residual estimate remains in the regular oscillatory regime. A positive exponent estimated from a time series does not necessarily imply a positive asymptotic Lyapunov exponent, since time-series methods can preferentially capture locally expanding dynamics even in a globally nonchaotic system~\cite{shuai2001positive}. In the present system, such finite-time effects may additionally be influenced by
the nonlinear optomechanical dynamics underlying the multimode oscillation. We therefore focus on the pronounced enhancement of the estimated LLE in the broadband regime, rather than its absolute sign alone.

\end{document}